\documentclass[aps,prl,showpacs,floatfix,reprint,superscriptaddress]{revtex4-1}

\pdfoutput=1

\usepackage{graphicx,color}
\usepackage{amssymb,amsthm,amsmath,hyperref,epsfig,wrapfig}

\begin{document}

\title{Orbital and Pauli limiting effects in heavily doped Ba$_{1-x}$K$_x$Fe$_2$As$_2$}

\author{Shuai Zhang}
\affiliation{Department of Physics, Kent State University, Kent, Ohio 44242, USA} 
\author{Y. P. Singh}
\affiliation{Department of Physics, Kent State University, Kent, Ohio 44242, USA} 
\author{X. Y. Huang}
\affiliation{Department of Physics, Kent State University, Kent, Ohio 44242, USA} 
\author{X. J. Chen}
\affiliation{Center for High Pressure Science and Technology Advanced Research, Shanghai 201203, China}
\author{M. Dzero}
\affiliation{Department of Physics, Kent State University, Kent, Ohio 44242, USA} 
\affiliation{Max Planck Institute for Physics of Complex Systems, 01187 Dresden, Germany}
\author{C. C. Almasan}
\affiliation{Department of Physics, Kent State University, Kent, Ohio 44242, USA} 

\date{\today}

\begin{abstract}
We investigated the thermodynamic properties of the Fe-based lightly disordered superconductor Ba$_{0.05}$K$_{0.95}$Fe$_2$As$_2$ in external magnetic field $H$ applied along the FeAs layers ($H||ab$ planes). The superconducting (SC) transition temperature for this doping level is $T_c = 6.6$ K. Our analysis of the specific heat $C(T,H)$ measured for $T<T_c$ implies a sign change of the superconducting order parameter across different Fermi pockets. We provide experimental evidence for the 
three components superconducting order parameter. We find that all three components have values which are comparable with the previously reported ones for the stochiometric compound KFe$_2$As$_2$.  Our data for $C(T,H)$ and resistivity 
$\rho(T, H)$ can be interpreted in favor of the dominant orbital contribution to the pair-breaking mechanism at low fields, while Pauli limiting effect dominates at high fields, giving rise to a gapless superconducting state with only the leading non-zero gap. 
\end{abstract}


\maketitle

\section{Introduction}
Materials with multiple Fermi surfaces may host unconventional superconductivity even for the case when one considers electron-phonon and Coulomb interactions only \cite{Agtenberg1999}. Since the interactions between the electrons from the same band as well as between the bands are fixed by the symmetry of the underlying orbitals, the resulting symmetry of a superconducting order parameter does not necessarily have to be of the conventional $s$-wave type. While more exotic pairing machanisms such as spin fluctuation exchange mechanism may naturally lead to the realization of exotic superconductivity in multiband materials with $d$- or $f$-electronic orbitals, it is generally expected that the microscopic structure of the superconducting order parameter remains universal under pressure or doping. 

The discovery of the `122' family of iron-based superconductors \cite{Rotter2008} has provided an example of unconventional superconductivity with a non-universal gap structure \cite{LeePRL2009,Tanatar2010,Maiti2012,MaitiPRB2013,GaraudPRL2014,Kang2014}.
In particular, in the hole-doped Ba$_{1-x}$K$_x$Fe$_2$As$_2$ high-$T_c$ superconductor, the superconducting order 
parameter changes
its symmetry with increasing doping level from lightly doped to over-doped regimes: at optimal hole doping ($x = 0.4$), the SC gap symmetry is of $s^{\pm}$ type \cite{Ding2008,Shimojima2011} with sign change between the isotropic fully-gapped hole Fermi pockets at the $\Gamma$ point and electron Fermi pocket  at the \textit{M} point in one Fe Brillouin zone (BZ), while the nodes in the gap appear in heavily doped alloys as $x\to 1$\cite{HashimotoPRB2010,OkazakiScience2012,Tafti2013,WatanabePRB2014}. This property appears to be universal as it is observed in other iron-based superconductor compunds \cite{SatoPRL2009,MazinNature2010,YinNature2011} and is most likely driven by near degeneracy between the superconducting states with different gap structure \cite{Zhang2009,Graser2010,Bernevig2011,Maiti2012,MaitiPRB2013,Fernandes2013}.  

Of a particular interest are the heavily overdoped Ba$_{1-x}$K$_x$Fe$_2$As$_2$ compounds. An important feature of these materials is that the electron pockets near the $M$ points are almost completely gone via the topological Lifshitz transition at $x\approx0.9$, while another hole band appears off-center relative to the BZ corner \cite{SatoPRL2009,HashimotoPRB2010,SuffianPRL2014}. Thus, the electronic properties are governed by the three hole pockets near the $\Gamma$ point. 
Importantly, the superconducting critical temperature does not vary significantly for doping levels just below or above the Lifshitz transition  \cite{XuPRB2013}. Furthermore, laser-excited angle-resolved photoemission spectroscopy measurements by Okazaki et al. \cite{OkazakiScience2012} have revealed a highly unusual gap structure in KFe$_2$As$_2$ with octet-line nodes on the middle hole Fermi sheet and nodeless gaps on inner and outer Fermi surfaces. Interestingly, recent specific heat measurements indicate that the nodes on the middle Fermi sheet are most likely accidental \cite{Hardy2014}. Subsequent measurements \cite{Tafti2013} of the changes in the superconducting critical temperature with pressure $P$ in KFe$_2$As$_2$ have revealed an evolution from a $d$-wave gap for $P<P_c$ to a $s^{\pm}$ gap for $P>P_c$. 
Lastly, the temperature dependence of thermal conductivity $\kappa(T)$ and its dependence on current direction and magnetic field also seem to be in agreement with nodal superconducting state  \cite{ReidPRL2012,ReidSUST2012}. 

In this paper we focus on the bulk properties of specific heat in single crystals of  Ba$_{0.05}$K$_{0.95}$Fe$_2$As$_2$. In particular, based on the measurements of specific heat performed in the $H||ab$ plane geometry (fixed direction of the magnetic field) we provide experimental evidence for the existence of a three-component superconducting order parameter. Importantly, we find that at least two components have opposite sign. Experimental evidence on whether one of the order parameter components possesses nodes is inconclusive. We also discuss the competition between spin and orbital pair breaking effects due to the generation of the Josephson vortices formed with a core running parallel to the FeAs planes. Specifically, our analysis of the field dependence of the Sommerfeld coefficient shows that when magnetic field exceeds $~ 4$T, the system is in gapless superconducting state with only one non-zero order parameter component. 

\section{EXPERIMENTAL DETAILS}
Single crystals of Ba$_{0.05}$K$_{0.95}$Fe$_2$As$_2$ with typical dimensions $1.3\times0.4\times0.2$ mm$^3$ were grown using the K-As flux method  \cite{KunihiroJPSJ2010}.  The doping level of the crystals used in this study was determined based on previously reported $T_c-x$ phase diagram \cite{RotterACIE2008,AvciPRB2012,StoreyPRB2013}. The specific heat $C$ and electrical resistivity $\rho$ were measured as a function of temperature $T$ and magnetic field $H$ with $H || c$ axis and $H || ab$ planes. $C(T,H)$ was measured using a relaxation technique in a field cooled manner by decreasing the temperature down to 0.5 K for $H || ab$ planes and to 2 K for $H || c$ axis. The resistivity measurements were carried out on samples with an {\it ac} electrical current $I_{ab}$ flowing in \textit{ab} plane and with $H$ always perpendicular to $I_{ab}$. 

\section{RESULTS AND DISCUSSION}
\subsection{Temperature dependence of the specific heat} 
Measurements of the specific heat provide an effective way to reveal the structure 
and the symmetry of the SC gap. 
Importantly, the magnetic-field dependence of the zero-temperature electronic specific heat coefficient ($\gamma$) measured in the mixed state with $H||c$ captures the details of nodal character of the gap symmetry. e.g., Volovik's theory predicts a $\sqrt H$ dependence of $\gamma$ in the case of a $d$-wave order parameter \cite{VolovikLett1993}. However, the presence of  $\sqrt H$ does not guarantee the nodal character of the gap and can arise due to other reasons. For example, a recent theoretical study of the $s^{\pm}$-wave state in iron-based superconductors shows that a change in the sign of the SC order parameter across different Fermi pockets also leads to the $\sqrt H$ dependence of $\gamma$ \cite{Bang2010}. 

\begin{figure}
\includegraphics[width=1\linewidth]{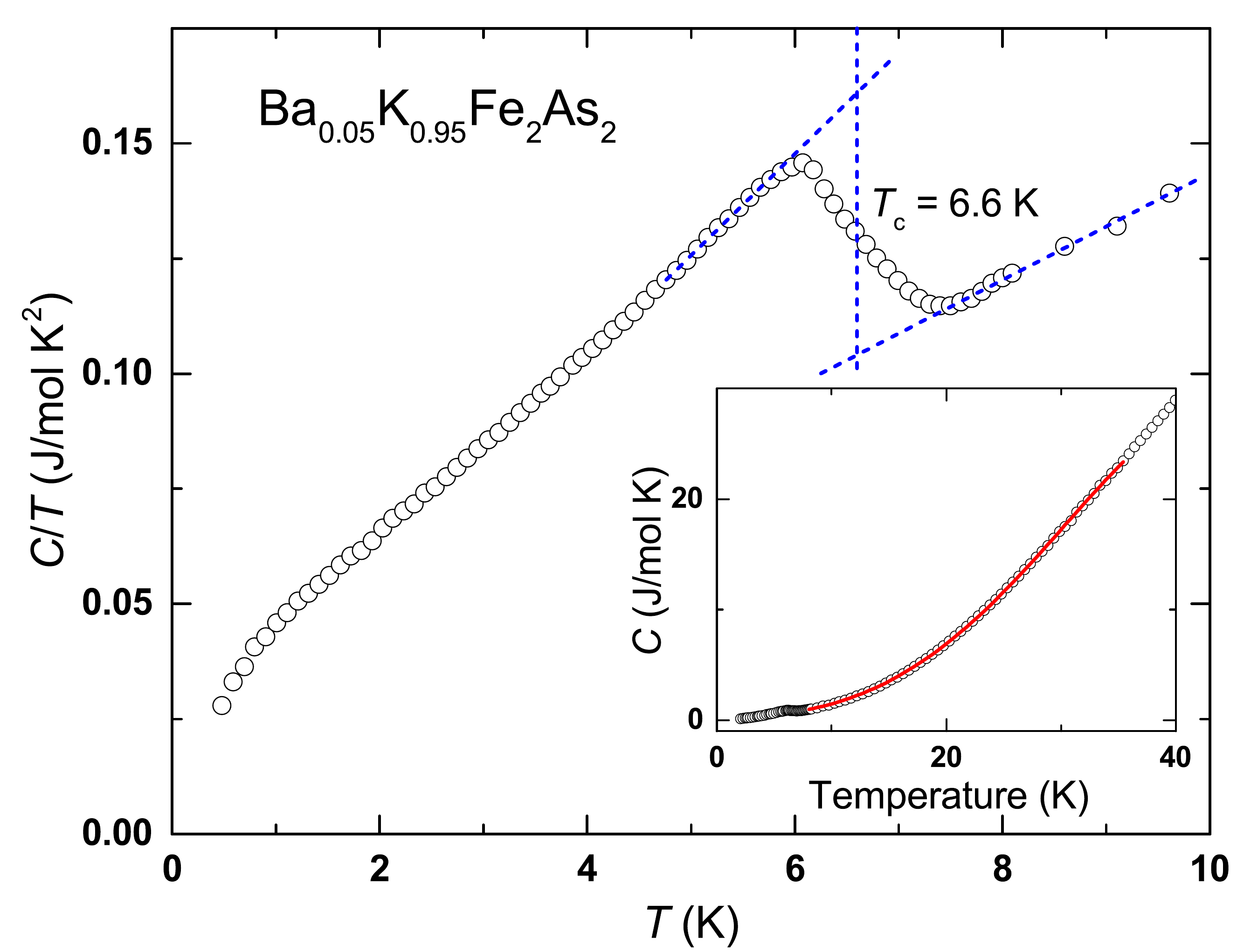}%
\caption{\label{f1} (Color online) Temperature dependence of $C/ T$ for Ba$_{0.05}$K$_{0.95}$Fe$_2$As$_2$ measured in the temperature range 0.5 K$\leq T \leq $ 10 K. The dotted line represents the isoentropic construction used to determine ${T_c}$. The inset shows a fit of normal state specific heat $C$ data using an expression $C$ = $\gamma_n T$+$\beta T^3$ (8 K$\leq T \leq $ 35 K), which is performed to determine the phonon contribution to the measured specific heat.}
\end{figure}

On Figure~\ref{f1} we show the $C/T$ data vs $T$ for Ba$_{0.05}$K$_{0.95}$Fe$_2$As$_2$ measured in zero field. The superconducting transition region has a width of about 1 K, which is typical for the overdoped samples of this family  \cite{HafiezPRB2012,SergeyPRB2012,SergeyPRB2013,StoreyPRB2013}. For the unambiguous determination of the thermodynamic superconducting transition temperature $T_c$ we use the \emph {isoentropic} construction (dotted blue line in the main panel), i.e., we choose $T_c$ such as the entropy around the transition is conserved.

Notice that the data of Fig.~\ref{f1} show a clear shoulder at low temperatures ($T< 2$ K). A similar feature has been observed in other multiband superconductors and it is indicative of the presence of additional SC gap(s) appearing at low temperatures (see e.g., \cite{Bouquet2001}). In fact, previous publications on  KFe$_2$As$_2$ have shown that there are at least three gaps present in this system  \cite{OkazakiScience2012, KawanoPRB2011}. 

Also notice that there is no low temperature upturn (Schottky nuclear contribution) in any of the specific heat data measured to temperatures as low as 0.5 K and fields $H||ab$ up to 14 T. Thus, the measured specific heat can be expressed in terms of electronic and lattice contributions as $C \equiv C_{\rm e} + C_{\rm ph}$. A fit of the normal state (8 K$\leq T \leq $ 35 K) specific heat data with $C$ = $\gamma_n T$+$\beta T^3$ gives $\gamma_n = 80$ mJ/mol$\cdot$K$^2$ and $\beta = 0.79$ mJ/mol$\cdot$K$^4$. The inset to Fig.~\ref{f1} shows the quality of this fit. The obtained $\gamma_n$ value is  comparable to the values of 69-103 mJ/mol K$^2$ for KFe$_2$As$_2$\cite{HidetoJPSJ2009,HafiezPRB2012,StoreyPRB2013,HardyPRL2013,HardyJPSJ2014} and it is slightly higher than the range 50-60 mJ/mol K$^2$ of the optimally doped ($x= 0.4$) samples  \cite{KantPRB2010,PopPRL2010}. Based on the $\beta$ value, we obtained a Debye temperature $\theta_D=230$ K.
To determine the electronic contribution to specific heat, we subtracted  the lattice contribution $\beta T^3$ from the measured specific heat. 

\begin{figure}
\includegraphics[width=1.0\linewidth]{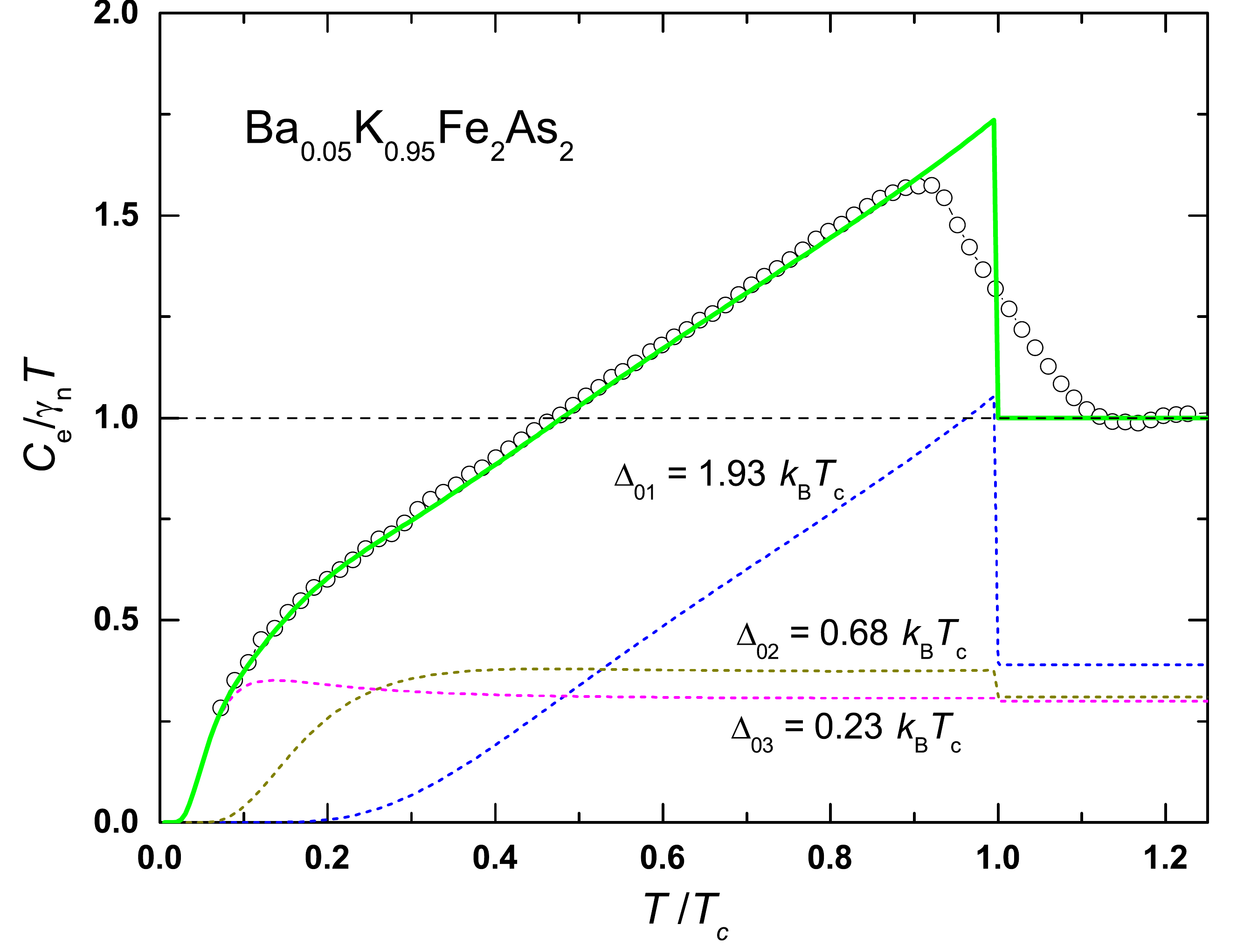}%
\caption{\label{f2} (Color online) Plot of the electronic specific heat normalized to its normal state value $C_{\rm e}/\gamma_n T$ vs normalized temperature $T/T_c$ (open circle). The solid green line is a fit of the data using the BCS model generalized to three bands (see text). The three  dashed lines show the contribution of the three individual superconducting gaps.  }
\end{figure}

To extract the values of the superconducting gaps from the specific heat data of Ba$_{0.05}$K$_{0.95}$Fe$_2$As$_2$, we plot the electronic specific heat normalized to its normal-state value vs $T/T_c$  (Fig. ~\ref{f2}). This plot shows that the specific heat decreases linearly with decreasing temperature over the range  0.2 $T_c\leq T \leq 0.85$ $T_c$ followed by a sharp drop around $T= 0.15$ $T_c$.  This overall $T$ dependence is very similar to that of KFe$_2$As$_2$  \cite{HafiezPRB2012,HardyPRL2013,HardyJPSJ2014}. 
The solid green line on the figure is a fit of these data using the BCS model generalized to three bands \cite{NagamatsuNature2001,BouquetEPL2001} and, for the time being, ignoring both the fact that the intermediate gap may be nodal \cite{OkazakiScience2012} and the effect of disorder, which might induce in-gap states. The details of the procedure used to determine the three gaps by fitting these data with Eq. 2 are discussed in the next paragraph. Notice the excellent fit of the data with this expression. The gap values are $\Delta_{0i}$ = 1.93, 0.68, and 0.23 $k_{\rm B}T_c$. The other fitting parameters are $r_1 = 0.39$, $r_2=0.31$, and $r_3=0.30$. Separate plots of the contribution of each gap (dashed lines on the figure) show that the smallest gap dominates at low $T$, while the largest gap gives the linear $T$ dependence. 

We remind the reader that within the BCS theory of superconductivity, for the case of an isotropic superconductor, the temperature dependence of the order parameter is determined self-consistently from the following results in the two limiting cases:
$\Delta(T)\propto\sqrt{(T_c-T)/T_c}$ at $T\sim T_c$ and $\Delta(T)\approx\Delta_0[1-\sqrt{2\pi T/\Delta_0}\exp(-\Delta_0/k_BT)]$ for $T\ll T_c$. Consequently, the temperature dependence of the entropy $S$ is given by \cite{BCS1959}:
\begin{equation}
\begin{split}
{S}= -\frac{6\gamma_n}{\pi^2k_B}\int_0^\infty \! [f_\epsilon \mathrm{ln}f_\epsilon + (1-f_\epsilon)\mathrm{ln}(1-f_\epsilon)]{d}\epsilon, 
\end{split}  
\end{equation}
where integration is over single particle energies $\epsilon$, and 
$f_\epsilon=[\exp(\sqrt{\epsilon^2+\Delta^2(T)}/k_BT)+1]^{-1}$ is the Fermi-Dirac distribution function. The electronic specific heat is calculated using the standard thermodynamic expression ${C_e} = T({\partial S}/{\partial T})_V$.  
Based on this, we used a superposition of three different SC gaps $\Delta_{0i}$ (\textit{i} = 1, 2, 3), hence three different $C_e$ with different relative contribution weights $r_i$:  
\begin{equation}\label{FitCe}
C_e(T) = \sum\limits_{i=1}^3r_iC_e(\Delta_{0i}, T).  
\end{equation}

Next, we compare the gap values obtained from our analysis with the previously reported values for KFe$_2$As$_2$ \cite{OkazakiScience2012,KawanoPRB2011}. Specifically, Table 1 lists the gap values from ARPES [\onlinecite{OkazakiScience2012}], $C(T)$ of present study, and small angle neutron scattering (SANS) \cite{KawanoPRB2011}. Notice that in all three cases there is a leading gap that is significantly larger than the other two gaps. Furthermore, our analysis and the SANS study give almost the same values for the three gaps, while the ratios of consecutive gap values (e.g., $ \Delta_{01}/ \Delta_{02}$ and $ \Delta_{02}/ \Delta_{03}$) are around 2.8 for both ARPES and our $C(T)$ study. We also note that the ARPES results have shown that the three components of the superconducting order parameter correspond to the three hole pockets near the $\Gamma$ point, with a nodal intermediate gap and a nodeless leading gap that is significantly larger than the other two. A comparison of all these findings implies that the superconducting order parameter of Ba$_{0.05}$K$_{0.95}$Fe$_2$As$_2$ has three different components, centered around the $\Gamma$ point with the values given above and with the middle gap nodal, but, as we will show later, most likely they are not protected by symmetry.  

\begin{table}
\small
 \caption{The superconducting gap values of Ba$_{1-x}$K$_x$Fe$_2$As$_2$ obtained in this study for the $x=0.95$ sample are compared with those reported in ARPES and SANS for the \textit{x} = 1 sample. The gaps are given in units of $k_{\rm B}T_c$.}
 \label{t1}
\begin{center}
\begin{tabular}{ccccc}
\hline\hline ARPES&&C(T)&&SANS\\ 
\hline
 $x = 1$ Ref. [\onlinecite{OkazakiScience2012}]&&$x = 0.95$&&$x = 1$ Ref. [\onlinecite{KawanoPRB2011}] \\
\hline 3.8&inner&1.93&&1.77\\
1.4&middle&0.68&&0.72\\
0.5&outer&0.23&&0.21\\
\hline\hline
\end{tabular} 
\end{center}
\end{table}

\subsection{Magnetic field dependence of the specific heat}

\begin{figure}
\includegraphics[width=1.0 \linewidth]{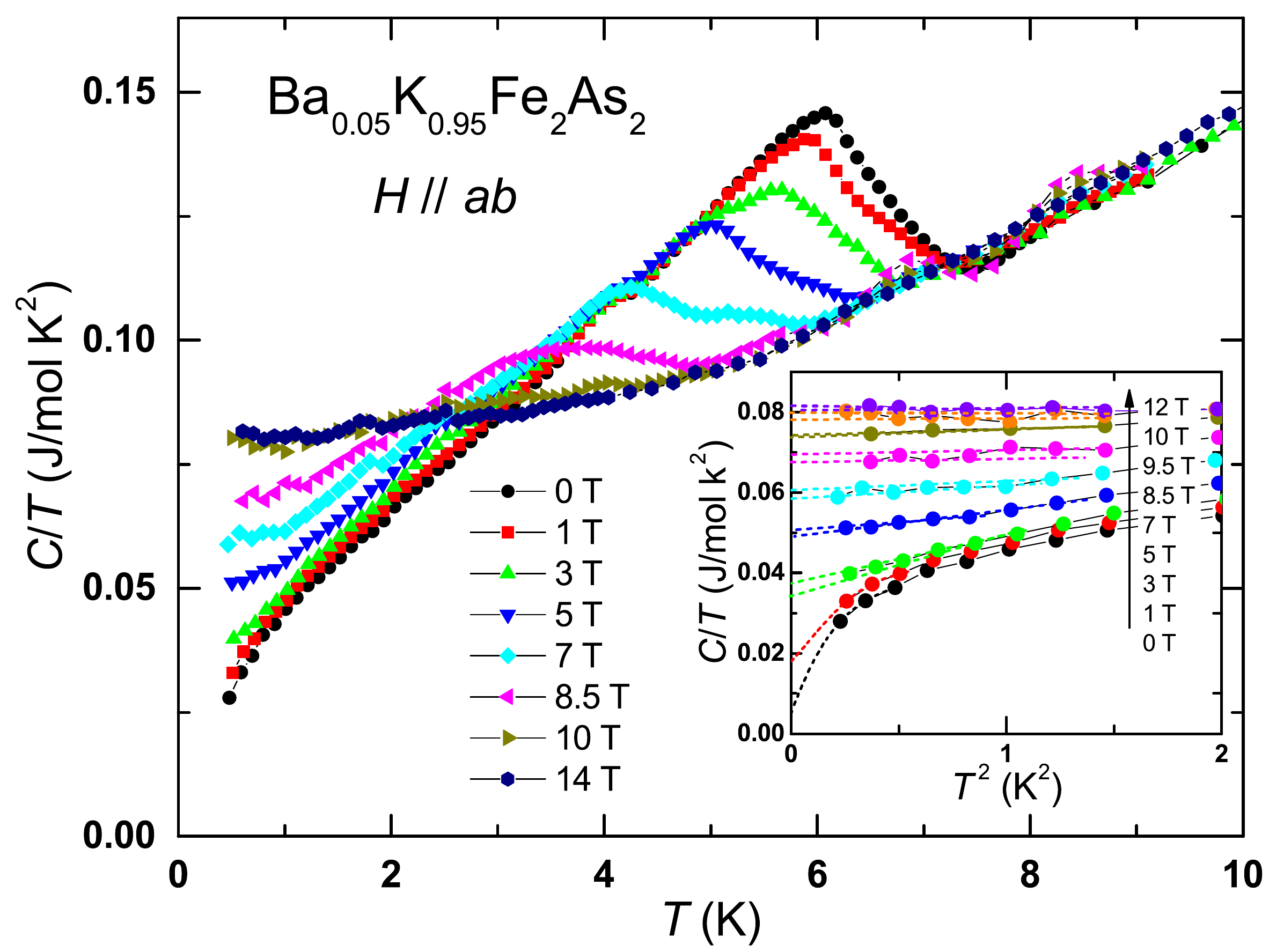}
\caption{\label{f3}(Color online) Temperature $T$ dependence of the specific heat  $C/T$ of Ba$_{0.05}$K$_{0.95}$Fe$_2$As$_2$ measured for 0.5 K $\leqslant$ T $\leqslant$ 10 K and magnetic fields $0 \leq H \leq 14$ T with $H||ab$ planes.  Inset: $C/T$ as a function of $T^2$ measured at low temperatures. An extrapolation (dashed lines) of these data at $T=0$ gives the zero-temperature Sommerfeld coefficient $\gamma$. Only selected fields are shown on both figures.}
\end{figure}

Next, we show the results of the magnetic field dependence of the specific heat for $H||ab$ planes. 
Figure~\ref{f3} shows the temperature dependence of $C/T$ for different $H$ values for the Ba$_{0.05}$K$_{0.95}$Fe$_2$As$_2$ sample. The low temperature data (see inset to Fig.~\ref{f3}) clearly show that the behavior of $C/T$ vs $T$ is very different at low and high magnetic fields: a shoulder in $C/T$ is observed at low fields, while it varies quadratically in $T$ at high fields. As discussed before, the shoulder in $C/T$ is typical of multiband superconductors, with additional gap(s) being present at low $T$ and $H$. The quadratic temperature dependence at high fields is due to the lattice contribution to the specific heat and, as such, it is of no great interest here. 

\begin{figure}
\includegraphics[width=1.0\linewidth]{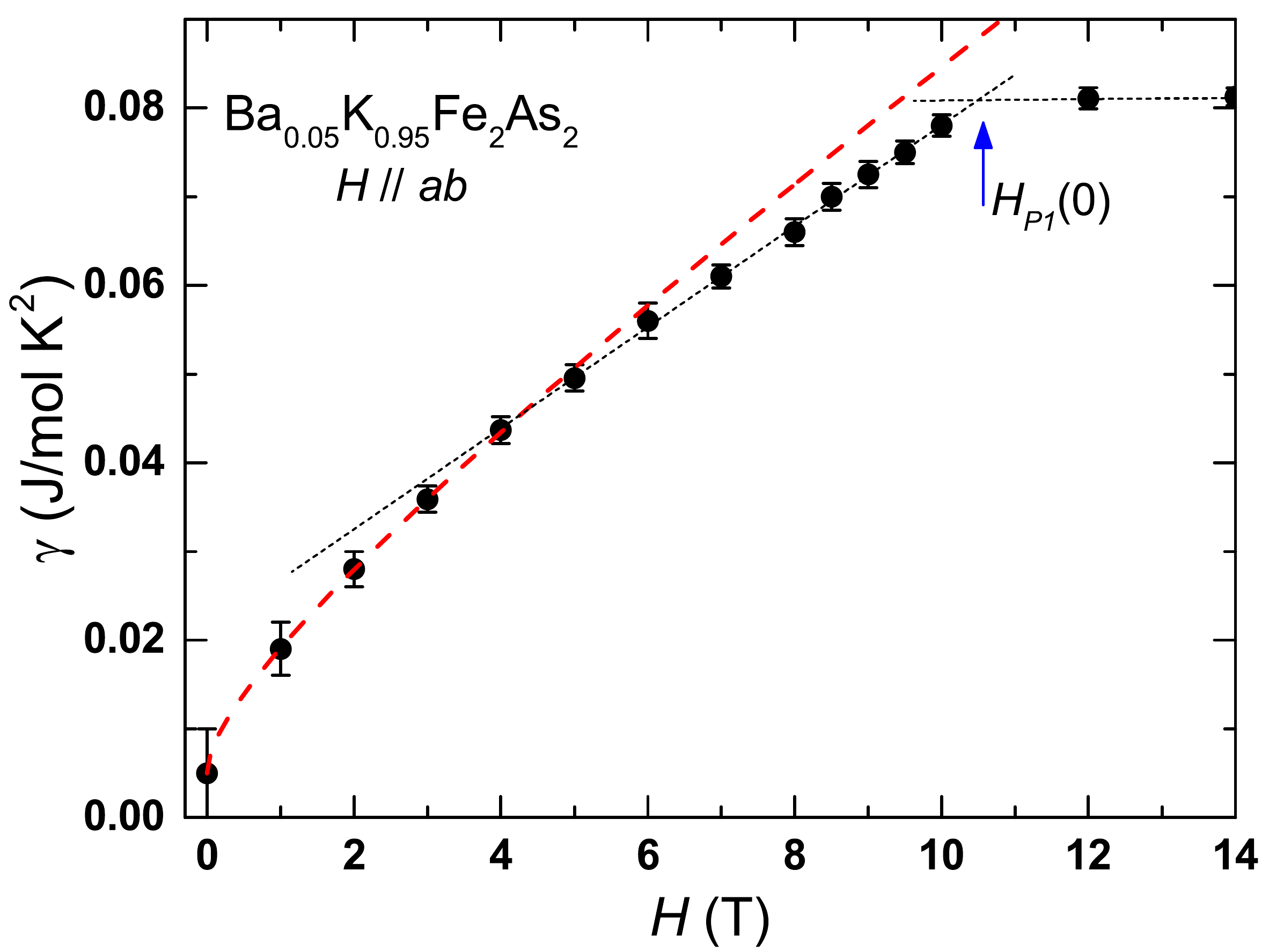}%
\caption{\label{fig4} (Color online) Magnetic field $H$ dependence of the zero-temperature Sommerfeld coefficient $\gamma$ of Ba$_{0.05}$K$_{0.95}$Fe$_2$As$_2$ for $H || ab$ planes. The red dotted line is a fit of the low $H$ data with Eq. (3), while the black dotted line is a linear fit of the high $H$ data. }
\end{figure}

The field dependence of the zero-temperature Sommerfeld coefficient $\gamma$ for $H||ab$ planes, in principle, allows us to probe the relative sign between two components of the superconducting order parameter, assuming that the third component has been fully suppressed by the magnetic field. We obtained the magnetic-field dependence of $\gamma$ (Fig.~\ref{fig4}) by extrapolating the data shown in the inset to Fig.~\ref{f3} to zero temperature. 
Notice the excellent fit of the data for $H\leq$ 4 T (red dashed line) with
\begin{equation}\label{GammaFit}
\gamma = a\cdot\sqrt{H} + b\cdot H + c, 
\end{equation}
where $a$ = 0.009 J/(mol K$^2$T$^{1/2}$), $b$ = 0.005 J/(mol K$^2$ T) and $c$ = 0.005 J/(mol K$^2$). The non-zero value of $c$
suggests the presence of non-superconducting impurities that give this residual $\gamma$ contribution. The impurity amount is about 6\% since this value of $c$ is only 6\% of $\gamma_n$. Furthermore, the presence of this disorder is relevant for our analysis as it introduces states in the gap which contribute to $\gamma(H\to 0)$. 
As we discuss in detail in the next paragraph, the $\sqrt{H}$ dependence at low fields is the result of the change in the sign of the SC order parameter across different Fermi pockets \cite{Bang2010}  and also emphasizes the role of disorder-induced scattering. 

Physically, the $\sqrt H$ dependence of $\gamma$ for $H< 4$ T can be interpreted as follows. Let us formally consider the superconducting order parameter $\Delta({\mathbf k})$ which is defined everywhere in the Brillouin zone. Then, consider two Fermi sheets FS$_1$ and FS$_2$ on which the order parameter has opposite sign: $\Delta({\mathbf k}_{\rm FS_1})=|\Delta_{01}|$ and $\Delta({\mathbf k}_{\rm FS_2})=-|\Delta_{02}|$. Clearly, on an arbitrary line connecting the two Fermi sheets in momentum space there is a point ${\mathbf k}_0$ where the order paramete vanishes, $\Delta({\mathbf k}_0)=0$. 
In the presence of disorder, single particle scattering can involve states with momentum transfers that involve parts of the Brillouin zone between the different Fermi sheets where the order parameter vanishes. These scattering processes effectively mimic the presence of nodes in the superconducting order parameter, resulting in the $\sqrt H$ dependence of $\gamma$ at lower fields, instead of the expected linear-in-$H$ behavior \cite{Bang2010}. 
Also, note that, when the Fermi sheets are sufficiently close to each other, as is the case in Ba$_{0.05}$K$_{0.95}$Fe$_2$As$_2$, the disorder does not necessarily have to be strong, unlike in the case when the relevant Fermi sheets are significantly separated in the BZ, say, one is around $\Gamma$ point, while another one is around $M$ point. Therefore, a 6\% disorder, as observed for this doping, seems to be sufficient to give the $\sqrt H$ dependence in $\gamma(H)$, i.e., to mimic the presence of nodes. We also note that, although in general, for a fixed magnetic field parallel to the FeAs planes, specific heat experiment cannot directly probe the nodal structure for each component of the order parameter, it can still probe whether the current lines pass through parts of the Fermi sheets with different or same sign of the corresponding pairing component and measure the relative sign of the order parameter components, provided the disorder-induced scattering is present in the system.   Moreover, the above theoretical approach based on the Volovik's effect implies that the orbital effect is the dominant pair-braking mechanism. As we will show later (Fig. ~\ref{f5}), in Ba$_{0.05}$K$_{0.95}$Fe$_2$As$_2$ with $H || ab$ planes, the orbital effect dominates at low fields, while Zeeman effect is the dominant pair breaker at high fields.

This result when compared with the other studies, including the most recent one by Kim et al.\cite{Kim2015}, can in principle be interpreted as evidence that the nodal structure of the intermediate component of the order parameter is most likely not symmetry protected, i.e., the nodes are accidental. In fact, recent theoretical studies \cite{Maiti2012}, specifically addressing the structure of the superconducting gap in KFe$_2$As$_2$, show that the states with and without accidental nodes are very close to degeneracy. 

We determine the Pauli limiting field $H_{P2}(0)$ for  the $\Delta_{02} = 0.68$ $k_{\rm B}T_c$ gap as follows. It is well known that there is a linear correlation  between $H_{c2}(0)$  and the superconducting gap $\Delta_0$  if the Pauli paramagnetic limit is the dominant pair-breaking mechanism, which is the case when $H||ab$ planes. Thus, we get $H_{P2}(0)=3.7$ T for the gap $\Delta_{02} = 0.68$ $k_{\rm B}T_c$ if we take  $H_{P1}(0) = 10.5$ T (see Fig.~\ref{fig4}) for the largest gap $\Delta_{01} = 1.93$ $ k_{\rm B}T_c$. The fact that this value is very close to 4 T, where there is the crossover between the sublinear and the linear in $H$ dependence of $\gamma$ (see Fig. \ref{fig4}), further shows that the order parameter has only one non-zero component for $H>H_{P2}$, so that the system is in a gapless superconducting state \cite{Barzykin2007}. A detailed study of this state will be the focus of future work. 

\subsection{H-T phase diagram}
\begin{figure}
\includegraphics[width=1.0 \linewidth]{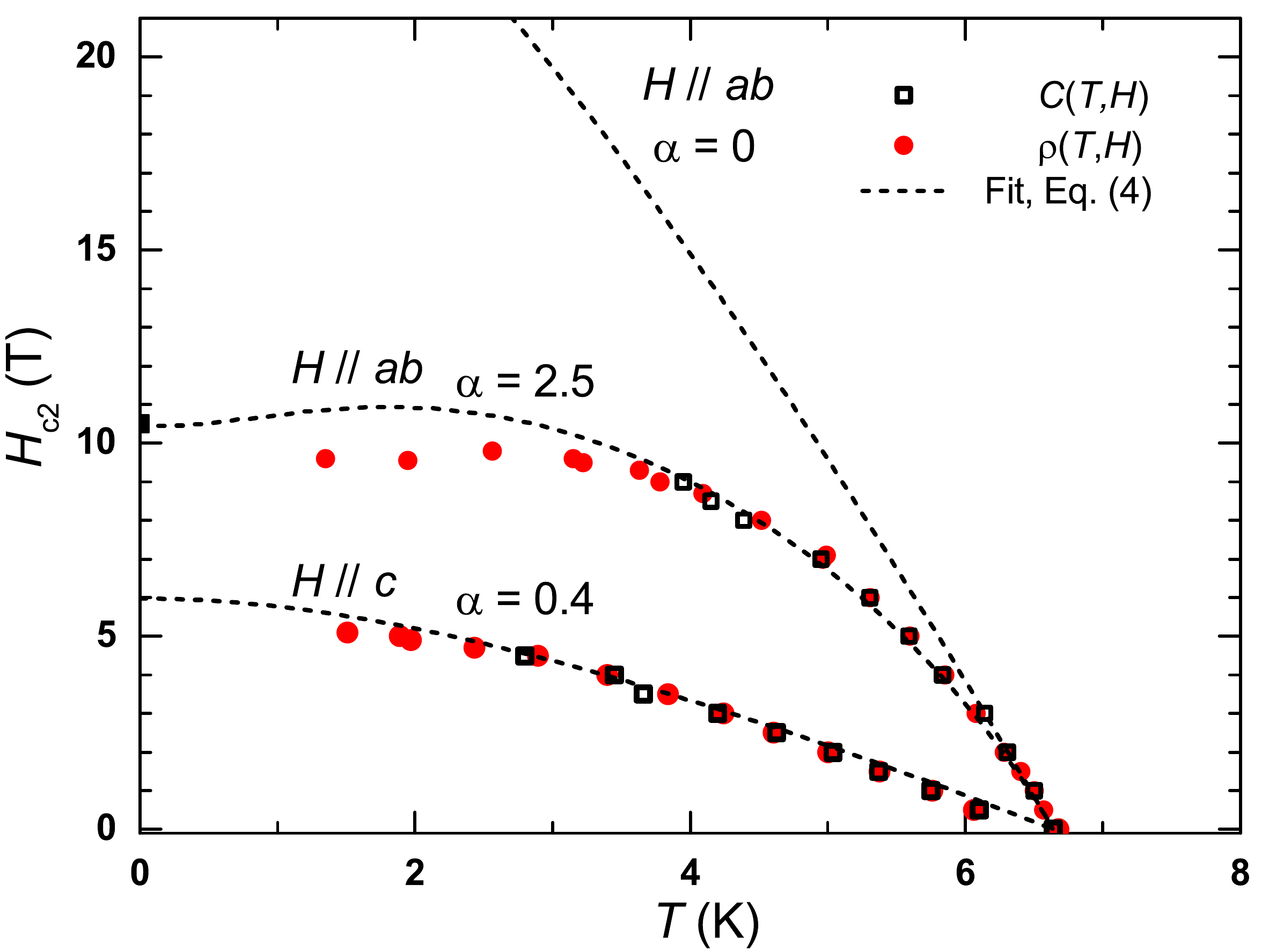}
\caption{\label{f5}(Color online) The upper critical field $H_{c2}$ vs $T$ phase diagram of Ba$_{0.05}$K$_{0.95}$Fe$_2$As$_2$. The dashed lines are fits of the data with the WHHM model ignoring the spin-orbit coupling contribution, i.e., $\lambda=0$ (see discussion for Eq. (\ref{FitWHHM}) for details).}
\end{figure}

For magnetic fields $H||ab$ planes, one generally expects a competition between the pair breaking processes due to Zeeman and orbital effects. 
However, as we discussed above, our data for $\gamma(H)$ (Fig. \ref{fig4}) clearly shows a crossover behavior from orbital dominated regime at 
low fields ($H<4$ T) to Zeeman effect dominated regime at high fields, suggesting an interplay between the momentum and spin degrees of freedom. 
In iron-pnictide superconductors this problem is well defined thanks to the negligibly small magnitude of the spin-orbit coupling. The presence of such an interplay is somewhat surprising given that the field configuration $H|| ab$ should lead to the much more dominant role of the Pauli limiting effects down to small magnetic fields just as it happens in the  `115' heavy-fermion superconductors such as CeCoIn$_5$. 
In order to quantify the contributions the orbital and Pauli limiting 
effects, we study the temperature dependence of the upper critical field $H_{c2}^{||}(T)$ for $H|| ab$ planes.

The superconducting critical temperatures under different fields with $H || ab$ planes and $H || c$ axis were determined using the isoentropic method discuss earlier and shown in Fig. \ref{f1}. Since the SC transition region in $C(T,H)$ broadens with increasing $H$ (see Fig. \ref{f3}), it becomes considerably difficult to determine the thermodynamic $T_c(H)$ at high $H$ values from $C(T,H)$. Thus, we determined $T_c(H)$ from the resistivity $\rho(T,H)$ data as the temperature of zero resistivity. We used this criterion since $T_c$ determined using the isoentropic criterion in $C(T)$ corresponds to the temperature of zero resistivity $\rho(T)$ (data not shown). However, the difference between $T_c(H)$ determined from resistivity and specific heat increases with increasing field. This behavior is a result of the increase of the vortex contribution to dissipation with increasing $H$. Thus, the zero-resistivity $T_c(H)$ at higher field values is underestimating the thermodynamic $T_c(H)$. The resulting $H$-$T$ phase diagram obtained from $C(T,H)$ and $\rho(T,H)$ with both $H || c$ and $H || ab$ is shown in Fig.~\ref{f5} after correcting for the underestimation of the $T_c(H)$ determined from $\rho(T,H)$. 

The Werthamer-Helfand-Hohenberg-Maki (WHHM) model \cite{WHHPR1966,HelfandPR1966} has been known as a useful theoretical prescription to evaluate $H_{c2}(T)$. Importantly, the model  includes the effects of Pauli spin paramagnetism, spin-orbit scattering, and orbital diamagnetic effects. 
We fit out data with a single-band WHHM model using the following expression:
\begin{equation}\label{FitWHHM}
\begin{split}
\ln\frac{T_c}{T}&= \textrm{Re}\left\{\psi\left[\frac{1}{2}+(1+i\alpha)\frac{hT}{2T_c}\right]\right\}-\psi\left(\frac{1}{2}\right),
\end{split}
 \end{equation}
where $h = 4H_{c2}/[-\pi^2T_c(\mathrm{d}H_{c2}/\mathrm{d}T)_{T=T_c}]$, $\psi(z)$ is digamma function, and $\alpha$  is Maki parameter that accounts for the relative contribution of orbital effect and Pauli spin paramagnetism \cite{MakiP1964,MakiPR1966}. 
In Eq. (\ref{FitWHHM}) we ignore the effect of spin-orbit coupling. Fits of the data with Eq. (4) (dotted lines) give $\alpha = 2.5$ for $H||ab$ planes and $\alpha=0.4$ for $H||c$ axis. These values for the Maki parameter are comparable to those obtained for the stoichiometric KFe$_2$As$_2$ \cite{Zocco2013}. Notice the excellent  agreement between the $H_{c2}(T)$ data and these fits. They yield $H_{c2}^{||}(0)\approx10.5$ and $H_{c2}^{\perp}\approx6.0$ T for $H||ab$ and $H||c$, respectively. We note that the value $H_{c2}^{||}(0)\approx10.5$ T obtained from Eq. (4) is in very good agreement with $H_{P1}(0) = 10.5$ T obtained from $\gamma(H)$ shown on Fig. \ref{fig4}. 

The orbital limit at zero temperature and in the dirty limit is $H_{c2}^{\rm orb}(0)$ = -0.69$T_c$(d$H$/d$T$)$_{T = T_c}=27.5$ T ($T_c = 6.6$ K) for $H || ab$, much larger that $H_{c2}^{||}(0)\approx10.5$, and $H_{c2}^{\rm orb}(0)=6.4$ T for $H || c$, comparable with $H_{c2}^{\perp}\approx6.0$ T. We also calculated the orbital effect with $H||ab$ using Eq. (4) with $\alpha=0$, also shown in Fig.~\ref{fig4} as a dotted line. Notice that this latter result only reproduces the $H_{c2}(T)$ data at low fields ($H< 3.5$ T). In summary, all these results clearly show that, for $H||ab$, the orbital effects dominate at low fields ($H< 3.5$ T) and the Pauli paramagnetic limit dominates at high fields, while for $H || c$ the orbital limit is dominant over the whole field range.

One can also calculate the value of $\alpha$ in the dirty limit using the expression \cite{WHHPR1966}
\begin{eqnarray}
\alpha = 3e^2\hbar\gamma_n\rho_n /2m\pi k_{\rm B}^2 ,
 \end{eqnarray}  
where $\gamma_n$ and $\rho_n$ are the specific heat coefficient and resistivity in the normal state (at $T_c$), respectively. With the previously determined values of $\gamma_n\approx80$ mJ/mol K$^2$ and $\rho_n$ = 1.8 $\mu\Omega$cm for $H||ab$ planes, Eq. (5) gives $\alpha$ = 2.4$\pm$0.1. This value is in excellent agreement with the one obtained by fitting the data of Fig.~\ref{f5} for $H || ab$ with  Eq. (4). 

Another interesting feature of the data of Fig.~\ref{f5} for $H||ab$ is the re-entrance region at small temperatures. Usually, this type of behavior is a signature for an instability which ultimately results in a first order transition manifested in discontinuous changes in thermodynamic quantities such as magnetization or thermal expansion coefficient. Interestingly, such a transition has, indeed, been observed in KFe$_2$As$_2$ \cite{Zocco2013}. This suggests again that the physics of the heavily hole-doped compound Ba$_{0.05}$K$_{0.95}$Fe$_2$As$_2$ is very similar to the physics of KFe$_2$As$_2$. However, the detailed investigation of a first order transition in magnetic field goes beyond the scope of this paper. 

\section{SUMMARY}
In summary, we have studied the bulk properties of Ba$_{0.05}$K$_{0.95}$Fe$_2$As$_2$, which is between the Lifshitz transition (around $\sim$0.9) and the stochiometric KFe$_2$As$_2$. We show that the specific heat $C(T)$ can be fitted using the BCS theory generalized to the presence of three bands and extract the three gap values as $\Delta_{0i}$ = 1.93, 0.68, and 0.23 $k_{\rm B}T_c$. We also discussed the magnetic field dependence of the zero-temperature Sommerfeld coefficient $\gamma(H)$ with $H||ab$ and show that at least two of the three order parameter components have opposite signs. Our analysis of $
\gamma(H)$ reveals a gapless superconducting state at a magnetic field higher than the $H_{c2}\approx$ 4 T, while  the nodes previously reported in the intermediate component of the order parameter are most likely accidental. The $H_{c2}-T$ phase diagram is obtained from the resistivity and specific heat data and is analyzed using the WHHM theory. We found that the orbital effects provide the dominant pair-breaking mechanism at the low-$H$ regime with $H||ab$.  We observed remarkable
similarities between the thermodynamic and magnetic properties of Ba$_{0.05}$K$_{0.95}$Fe$_2$As$_2$ and stoichiometric compound KFe$_2$As$_2$.

\section*{ACKNOWLEDGMENTS}
This work has been supported by the National Science Foundation NSF DMR-1505826 at Kent State University. M.D. acknowledges financial support from KSU and MPI-PKS. 

 
\end{document}